\documentclass[10pt,pra,aps,twocolumn,superscriptaddress,floatfix,notitlepage,nofootinbib]{revtex4-2}
\usepackage[latin1]{inputenc}
\usepackage{hyperref}
\hypersetup{
    colorlinks=true,
    linkcolor=blue,
    filecolor=magenta,      
    urlcolor=cyan,
    citecolor=blue,
    pdftitle={Interacting Q-balls},
    pdfpagemode=FullScreen,
    }
\usepackage{mathrsfs,dsfont}
\usepackage{amsmath}
\usepackage{varwidth}
\usepackage{empheq}
\usepackage{braket}
\usepackage{amsfonts,natbib}
\usepackage{amssymb}
\usepackage{bm}
\usepackage[pdftex]{graphicx}
\usepackage[normalem]{ulem}

\usepackage{graphicx}
\begin{document}
\title{
Quantum Sensing with Joint Emitter-Fluorescence Measurements
}
\author{Yuliya Bilinskaya}
\email{yuliyab@kth.se}
\affiliation{KTH Royal Institute of Technology, Hannes Alfv\'{e}ns v\"{a}g 12, SE-106 91 Stockholm, Sweden}
\author{Sreenath K. Manikandan}
\email{skm@tifrh.res.in}
\affiliation{Tata Institute of Fundamental Research Hyderabad, 36/P, Gopanpally Village, Serilingampally Mandal, Hyderabad, Telangana 500046, India}

\newcommand{\yb}[1]{{\color[rgb]{0.,0.6,0.5}{[YB: #1]}}}
\newcommand{\ybadd}[1]{{\color[rgb]{0.,0.6,0.5}{#1}}}
\newcommand{\skm}[1]{{\color[rgb]{0.6,0.0,0.5}[SKM: #1]}}
\newcommand{\skmadd}[1]{{\color[rgb]{0.6,0.0,0.5}{#1}}}

\date{\today}
 \begin{abstract}
We present an analytically tractable model of a driven quantum harmonic emitter, such as an oscillating charged dipole, emitting radiation via resonance fluorescence.
With this model we are able to characterize the quantum mechanical correlations that are built up at early times between the drive, the resonant emitter, and its fluorescence.
We describe detection strategies that can reveal these quantum signatures in experiments by performing joint measurements on the quantum emitter and its fluorescence field. 
In particular, we show that simultaneous quantum measurements of a driven quantum emitter and its fluorescence can be used to probe the quantum noise of the driving field, relative to the maximally classical coherent state of the driving field, in short-time experiments.  
We conclude by discussing the applications to quantum sensing in quantum optical, quantum acoustic, and quantum gravitational scenarios of interest.

\end{abstract}
	\maketitle
\section{Introduction}
 Resonance fluorescence serves as a widely characterized case of secondary emission, where a resonantly driven external field can stimulate a quantum emitter to fluoresce, or in other words, to emit radiation quanta into a (vacuum) mode different from the driving field. 
 Experimental investigations probing resonance fluorescence of photons from atoms was crucial in the understanding of the non-classical features of light in tabletop experiments, famously in the early experiments by Mandel and Short \cite{short1983observation}, that verified the sub-Poissonian quantum statistics for light~\cite{kimble1977photon,mandel1979sub,singh1983antibunching,cook1981photon,carmichael1989photoelectron}. 
 Analogous processes involving emission of matter waves in ultracold atoms have also been studied in recent years \cite{kim2025super, stewart2020dynamics, krinner2018spontaneous, navarrete2011simulating, de2008matter}, demonstrating that the guiding principles of resonance fluorescence based on quantum optics are broadly applicable. 

Recent progress in circuit quantum electrodynamics has made it possible to observe resonant fluorescence emission time-continuously as well, with remarkable collection efficiencies \cite{lewalle2020measuring,campagne2016observing,jordan2016anatomy,campagne2014observing,campagne2016using,ficheux2018dynamics,albertinale2021detecting}. 
They allow one to go beyond the conventional quantum jump paradigm of photon number resolving measurements in fluorescence to the diffusive quantum measurements regime of phase-sensitive, and phase-preserving quantum measurements~\cite{lewalle2020measuring,jordan2016anatomy,campagne2016observing,karmakar2022stochastic}. 
With such generalized measurements, one could probe the gradual fluorescent decay of the atom through diffusive quantum trajectories that correspond to resonance fluorescence. 
On the fundamental side, such time-continuous detection strategies also enable analysing resonance fluorescence from a thermodynamic point of view~\cite{manikandan2019fluctuation,garrahan2010thermodynamics,szczygielski2013markovian,langemeyer2014energy,cuetara2015stochastic,donvil2018thermodynamics,elouard2020thermodynamics}. Emission scenarios driven by continuous quantum measurements have also been proposed, with applications extending to studying autonomous quantum clocks based on artificial atoms~\cite{manikandan2023autonomous,benny2025quantum,erker2017autonomous,singh2025quantum}. 

While the canonical formulations of probing resonance fluorescence emission involve two-level atoms as resonant emitters~\cite{milonni2013quantum,Weisskopf:1930au}, emission of radiation via resonance fluorescence is a specific example of secondary emission. Connecting back to earlier ideas in optics, Huygen's principle describes that propagating wavefronts obey a similar principle, where every point in the primary wavefront can be thought of as a source of a secondary wavefront. The corresponding spatial modes characterizing the primary or secondary sources can effectively be modelled as quantum harmonic oscillators. Harmonically oscillating quantum emitters such as oscillating charge dipoles or mass quadrupoles naturally offer a much simpler setting to capture the essential details of primary and secondary emission. They also find interesting applications in fields beyond quantum optics, such as in quantum acoustics where the physics of emitters revolve around quantum harmonic oscillators whose fundamental excitations are phonons~\cite{kim2025super, stewart2020dynamics, krinner2018spontaneous, navarrete2011simulating, de2008matter,stewart2017analysis}. Superradiant laser clocks also provide a notably relatable optical paradigm with important implications extending to precise-timekeeping~\cite{meiser2009prospects,bohnet2012steady}. 

 \begin{figure}[tb]
    \centering
\includegraphics[width=\columnwidth]{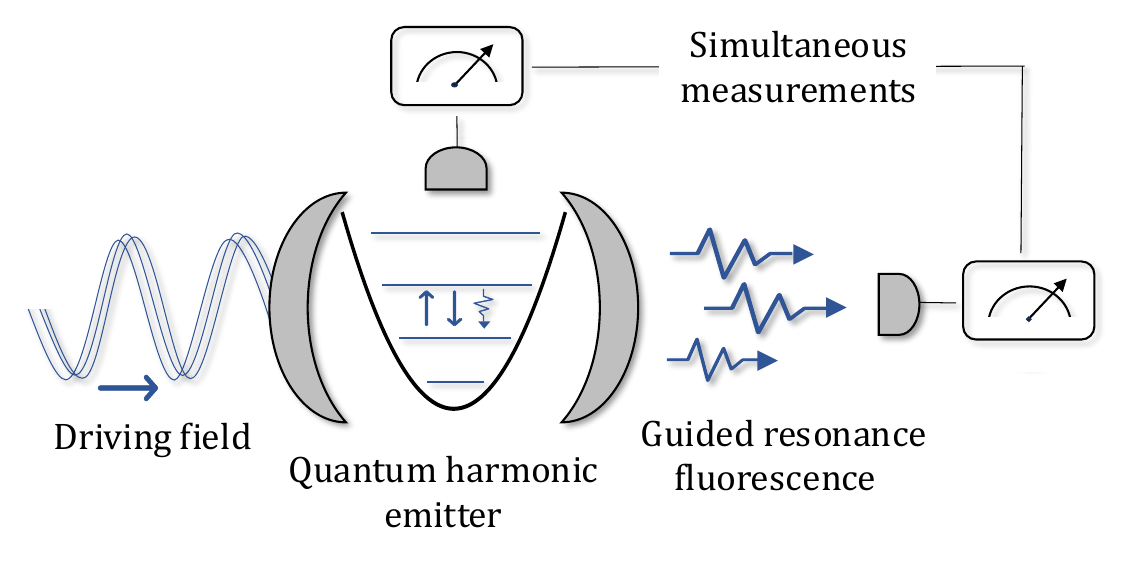}
    \caption{A quantum sensing scheme to probe the quantum noise characteristics of the driving field using simultaneous measurements of a driven quantum harmonic emitter and its fluorescence. This approach provides statistical null tests of classicality, relative to a coherent state of the driving field.}
    \label{fig:Model}
\end{figure}
The above aspects motivates us to consider an analytically tractable model for a driven, monitored quantum harmonic emitter. 
 Our key observation is that simultaneously monitoring the quantum emitter and its resonance fluorescence as shown in Fig.~\ref{fig:Model}, although they correspond to different physical modes, presents itself as an interesting avenue of incompatible observations in quantum mechanics. The reason is that the fluorescence emission, which is associated to the annihilation of quanta in the emitter, fundamentally disturbs the quantum state of the emitter. 
 This implies that observing resonance fluorescence emission simultaneously with the direct observations on the quantum emitter can give additional quantum information complementary  to the quantum information that can be acquired from direct quantum measurements on the emitter or its fluorescence alone.  

A major application of our findings is in the context of quantum sensing. 
The dynamics of the driven oscillator under a coherent driving field---a quantum field which behaves most classically---reveals that the emitter and its fluorescence are also coherent states, separable at all times. However this is no longer true for generic quantum mechanical states of the driving field.
The quantum features of the driving field in fact propagate into the emitter and its fluorescence field, which allows us to study the induced quantum features in secondary emission via correlation measurements on a quantum emitter and its fluorescence. 
In particular, by performing joint compatible measurements of position and momentum quadratures on the quantum emitter and its fluorescence, we are able to probe the the variance and covariance of the canonical position and momentum quadratures of the driving field. 
With this information we reconstruct the full quantum noise (covariance) matrix of the driving field, which offers an information theoretically complete description for Gaussian states. 

The covariance of counts is also discussed, which probes the quantum noise of the driving field in the number basis, and therefore offers additional insights.  Importantly, the correlation measurements we propose offer statistical ``null tests'' for classicality through joint measurements of a resonant emitter and its fluorescence in that it probes the quantum noise of the driving field relative to a coherent state, yielding the null outcome, zero, for a coherent state. In this aspect, our scheme of probing incompatible radiative processes in a single emitter are complementary to recent results on probing the quantum characteristics of radiation fields using multiple resonant detectors~\cite{manikandan_testing_2025,manikandan2025complementary,manikandan2025probing}, including the statistical null tests they offer~\cite{manikandan2026detector}, with implications to probing the non-classical aspects of the gravitational radiation field as well.
Broadly, our findings re-affirm the guiding principle of quantum optics that ``good quantum emitters are also good detectors", with important applications to probing the quantum characteristics of radiation fields in tabletop experiments.

\section{The Physical Model}
\label{section_Model}

We consider a resonantly driven quantum harmonic emitter, which de-excites via resonant fluorescent emission into another channel initialized in the vacuum.
The mode to which the quantum emitter emits into via resonance fluorescence is also modeled as a quantum harmonic mode which can accommodate an arbitrary number of quanta, photons for example.
Using the rotating wave approximation with the single-mode frequency of the driving field the interaction Hamiltonian in the interaction picture is defined as,
\begin{equation}
    \hat{H}_I(t) = \hbar \left( \sqrt{\gamma_0} \hat{a}^\dagger \hat{d}(t) + \sqrt{\gamma_s}  \hat{d}^\dagger(t) \hat{c}  + \text{h.c.}\right),
\end{equation}
where $\hat{a}$ is the mode of the driving (radiation) field, the $\hat{d}$ is the mode of the emitter and $\hat{c}$ is the mode of the resonance fluorescence and $\sqrt{\gamma_0}$, $\sqrt{\gamma_s}$ are the effective coupling strengths of the driving field with the emitter and the emitter with the resonant fluorescence respectively.
The interaction Hamiltonian retains only the resonant interaction terms, while the residual time-dependence in the $\hat{d}$ mode captures possible Markovian resetting of the emitter mode during de-excitation and re-excitation ensured by the commutator,
\begin{equation}
    [\hat{d}(t'), \hat{d}^\dagger(t'')]=\delta (t'-t'').
\end{equation}
The unitary time evolution operator of the model is $\hat{U}_I(t) = e^{-i \int_t^{t+\Delta t} \left( \sqrt{\gamma_0} \hat{a}^\dagger \hat{d}(t') + \sqrt{\gamma_s}  \hat{d}^\dagger(t') \hat{c}  + \text{h.c.}\right) dt'}$.
We can introduce the operators $\sqrt{\Delta t} \hat{b} \equiv \int_t^{t+\Delta t}  \hat{d}(t') dt'$ and $\sqrt{\Delta t} \hat{b}^\dagger \equiv \int_t^{t+\Delta t}  \hat{d}^\dagger(t') dt'$, satisfying the commutation relation $[\hat{b}, \hat{b}^\dagger]=1$, such that the time evolution operator becomes
\begin{equation}
    \hat{U}_I = e^{-i \sqrt{\Delta t} \left( \sqrt{\gamma_0} \hat{a}^\dagger \hat{b} + \sqrt{\gamma_s}  \hat{b}^\dagger \hat{c}  + \text{h.c.}\right)}.
    \label{U_I}
\end{equation}
In the single mode approximation we are working on,  $\gamma_0$ and $\gamma_s$ can be understood as spontaneous emission rate for the emitter to emit a quantum of the driving field and a quantum of fluorescence respectively. They can be same $\gamma_0 = \gamma_s$ if the driving field and the fluorescent field are of the same kind, optical fields for example in standard quantum optics scenarios. However, in more generalized settings, optical or gravitational excitations can lead to emission of a quanta of different kind, via matter waves for example~\cite{krinner2018spontaneous}, and we therefore label these two spontaneous emission rates with different subscripts. The time dependence $\sqrt{\Delta t}$, which arises from renormalizing the detector mode, also implies that the above interaction can be understood as an effective description that predicts probabilities increasing linearly with time, directly yielding Fermi Golden Rule probabilities~\cite{manikandan_testing_2025,lewalle2020measuring}. If the decay is via optical means, the rate $\gamma_s$ in our model is indeed the spontaneous emission rate predicted by the standard Wigner-Weisskopf theory~\cite{milonni2013quantum,Weisskopf:1930au}. Some physically relevant examples and the corresponding emission rates $\gamma_0,\gamma_s$ are discussed in Sec.~\ref{section_discussion}.

\subsection{Exact dynamics}

Initializing the driving field in the coherent state $\ket{\alpha}_a$, the emitter and the fluorescence field in their respective ground states $\ket{0}_b$ and $\ket{0}_c$, the dynamics is captured by the unitary time evolution operator from Eq.~\eqref{U_I} as, 
\begin{multline}
    \hat{U}_{I}\ket{\alpha}_a \ket{0}_b \ket{0}_c = \\e^{-i \sqrt{\Delta t} \left( \sqrt{\gamma_0} \hat{a}^\dagger \hat{b} + \sqrt{\gamma_s}  \hat{b}^\dagger \hat{c}  + \text{h.c.}\right)}\ket{\alpha}_a \ket{0}_b \ket{0}_c.
    \label{Time_evo_U_I}
\end{multline}
Rewriting the coherent state as a displaced vacuum state $\ket{\alpha}_a = e^{-|\alpha|^2/2} e^{\alpha \hat{a}^\dagger}\ket{0}_a$, Eq.~\eqref{Time_evo_U_I} can be written as
 \begin{equation}
     \hat{U}_{I}\ket{\alpha}_a \ket{0}_b \ket{0}_c = e^{-|\alpha|^2/2}e^{\alpha \hat{U}_I \hat{a}^\dagger \hat{U}_I^\dagger} \ket{0}_a \ket{0}_b \ket{0}_c.
     \label{full_state_evo}
 \end{equation}
To evaluate the expression for $\hat{U}_I \hat{a}^\dagger \hat{U}_I^\dagger$ we use the Hadamard's lemma: $e^{J} S e^{-J} = S + [J, S] + \frac{1}{2!} [J,[J, S]] +  \frac{1}{3!} [J,[J,[J, S]]] + ...$.
Setting $\gamma \equiv \gamma_0  + \gamma_s$, $S\equiv \hat{a}^\dagger$ and $J \equiv -i \sqrt{\Delta t}(\sqrt{\gamma_0}\hat{a}^\dagger \hat{b} + \sqrt{\gamma_0}\hat{a} \hat{b}^\dagger + \sqrt{\gamma_s}\hat{b}^\dagger \hat{c} + \sqrt{\gamma_s}\hat{b} \hat{c}^\dagger)$ we explicitly calculate the first few terms:

\noindent
$S = \hat{a}^\dagger$,

\noindent
$[J, S] = -i \sqrt{\Delta t}\sqrt{\gamma_0} \hat{b}^\dagger$,

\noindent
$[J,[J, S]] = (i \sqrt{\Delta t})^2(\gamma_0 \hat{a}^\dagger  + \sqrt{\gamma_s \gamma_0} \hat{c}^\dagger )$,

\noindent
$[J,[J,[J, S]]] =  -(i \sqrt{\Delta t})^3 \sqrt{\gamma_0} \gamma \hat{b}^\dagger$,

\noindent
$[J,[J,[J,[J, S]]]] = (i \sqrt{\Delta t})^4  \gamma (\gamma_0 \hat{a}^\dagger  + \sqrt{\gamma_s \gamma_0} \hat{c}^\dagger )$,

\noindent
$[J,[J,[J,[J,[J, S]]]] =  -(i \sqrt{\Delta t})^5 \sqrt{\gamma_0} \gamma^2  \hat{b}^\dagger$,

\noindent
$[J,[J,[J,[J,[J,[J,S]]]]]] = (i \sqrt{\Delta t})^6 \gamma^2 (\gamma_0 \hat{a}^\dagger  + \sqrt{\gamma_s \gamma_0} \hat{c}^\dagger )$,

\noindent
$[J,[J,[J,[J,[J,[J,[J, S]]]]]]] =  -(i \sqrt{\Delta t})^7 \sqrt{\gamma_0}\gamma^3 \hat{b}^\dagger$.

At this point we notice that there is a specific structure of the coefficients in front of each mode,  which for the driving field can be written as
\begin{equation}
\left(\frac{\gamma_0 }{\gamma} \sum_{k=1}^{k=\infty}  \frac{(-1)^k \left(\sqrt{\Delta t \gamma}\right)^{2k}}{(2k)!} + 1\right) \hat{a}^\dagger,
\label{Drive_coeff}
\end{equation}
the detector as 
\begin{equation}
-i \sqrt{\frac{\gamma_0}{\gamma}}  \sum_{k=0}^{k=\infty}  \frac{(-1)^k \left(\sqrt{\Delta t \gamma }\right)^{2k+1}}{(2k+1)!} \hat{b}^\dagger,
\label{Detector_coeff}
\end{equation}
and the fluorescence as
\begin{equation}
\frac{\sqrt{\gamma_0 \gamma_s}}{\gamma} \sum_{k=1}^{k=\infty}  \frac{(-1)^k \left(\sqrt{\Delta t \gamma}\right)^{2k}}{(2k)!}  \hat{c}^\dagger.
\label{Fluorescence_coeff}
\end{equation}
Rewriting the Eqs.~\eqref{Drive_coeff}-\eqref{Fluorescence_coeff} as trigonometric series gives 
\begin{multline}
\hat{U}_I \hat{a}^\dagger \hat{U}_I^\dagger = \\
\left[\frac{\gamma_0 }{\gamma} \text{cos}\left(\sqrt{\Delta t \gamma}\right) + \frac{\gamma_s }{\gamma} \right] \hat{a}^\dagger - 
i \sqrt{\frac{\gamma_0}{\gamma}} \text{sin}\Big(\sqrt{\Delta t \gamma } \Big) \hat{b}^\dagger + \\
\frac{\sqrt{\gamma_0 \gamma_s}}{\gamma} \left[\text{cos}\left(\sqrt{\Delta t \gamma }\right) - 1 \right] \hat{c}^\dagger.
\label{all_trig_series}
\end{multline}
Using Eq.~\eqref{all_trig_series} in Eq.~\eqref{full_state_evo} and rewriting the state as the product of coherent states we obtain the exact time evolution of a quantum emitter driven resonantly with a coherent field

\begin{multline}
     \hat{U}_{I}\ket{\alpha}_a \ket{0}_b \ket{0}_c =  \\
     \Big|\alpha \frac{1}{\gamma}\left[\gamma_0 \text{cos}\big(\sqrt{\Delta t \gamma}\big) + \gamma_s \right]\Big\rangle_a \otimes 
     \Big|-\alpha  i \sqrt{\frac{\gamma_0}{\gamma}} \text{sin}\big(\sqrt{\Delta t \gamma }\big)\Big\rangle_b \otimes \\
     \Big|\alpha  \frac{\sqrt{\gamma_0 \gamma_s}}{\gamma} \left[\text{cos}\big(\sqrt{\Delta t \gamma}\big)  - 1 \right]\Big\rangle_c.
     \label{exact_results}
\end{multline}
We note that this model could also be solved with an alternative method of normal modes presented in Appendix~\ref{appendix_normal_modes_solution}.
The solution in Eq.~\eqref{exact_results} is a product state of coherent states, which also demonstrates that the features of the driving field propagate via a scaling of amplitudes into the emitter and its fluorescence.
This simplicity allows to probe quantum features that would arise in the system once it deviates from the maximally classical, coherent state, behavior. %
In order to probe generic quantum features of the driving field, we represent an arbitrary quantum state of the driving field via the Sudarshan-Glauber P-representation~\cite{sudarshan1963equivalence,glauber1963coherent,Mandel_Wolf_1995},
\begin{equation}
    \rho_a (0)= \int d^2\alpha P(\alpha) |\alpha\rangle\langle\alpha|_a.
    \label{P_rep_rho}
\end{equation}
The properties of the P-function dictate the non-classical features of the state, for example, squeezed states necessarily take negative values for $P(\alpha)$ in some parts of the phase-space~\cite{gerry2006introductory}.

Using Eq.~\eqref{P_rep_rho} in Eq.~\eqref{exact_results} gives the state of the entire system for a generic state of the driving field,

\begin{multline}
     \hat{U}_{I}\left(\rho_a \otimes |0\rangle\langle0|_b \otimes |0\rangle\langle0|_c\right) \hat{U}_{I}^\dagger=  \int d^2\alpha P(\alpha)\\
      \big|\alpha \frac{1}{\gamma}\left[\gamma_0 \text{cos}\big(\sqrt{\Delta t \gamma}\big) + \gamma_s \right]\big\rangle \big\langle\alpha \frac{1}{\gamma}\left[\gamma_0 \text{cos}\big(\sqrt{\Delta t \gamma}\big) + \gamma_s \right]\big|_a \otimes \\
     \big|-\alpha  i \sqrt{\frac{\gamma_0}{\gamma}} \text{sin}\big(\sqrt{\Delta t \gamma }\big)\big\rangle \big\langle-\alpha  i \sqrt{\frac{\gamma_0}{\gamma}} \text{sin}\big(\sqrt{\Delta t \gamma }\big)\big|_b \otimes \\
     \big|\alpha  \frac{\sqrt{\gamma_0 \gamma_s}}{\gamma} \left[\text{cos}\big(\sqrt{\Delta t \gamma}\big)  - 1 \right]\big\rangle \big\langle\alpha  \frac{\sqrt{\gamma_0 \gamma_s}}{\gamma} \left[\text{cos}\big(\sqrt{\Delta t \gamma}\big)  - 1 \right]\big|_c.
     \label{exact_results_P_rep}
\end{multline}
Tracing out the driving field allows to obtain the reduced density matrix of the emitter and its fluorescence,
\begin{multline}
    \rho_{bc} =  \int d^2\alpha P(\alpha) \\
      \big|-\alpha  i \sqrt{\frac{\gamma_0}{\gamma}} \text{sin}\big(\sqrt{\Delta t \gamma }\big)\big\rangle \big\langle-\alpha  i \sqrt{\frac{\gamma_0}{\gamma}} \text{sin}\big(\sqrt{\Delta t \gamma }\big)\big|_b \otimes \\
     \big|\alpha  \frac{\sqrt{\gamma_0 \gamma_s}}{\gamma} \left[\text{cos}\big(\sqrt{\Delta t \gamma}\big)  - 1 \right]\big\rangle \big\langle\alpha  \frac{\sqrt{\gamma_0 \gamma_s}}{\gamma} \left[\text{cos}\big(\sqrt{\Delta t \gamma}\big)  - 1 \right]\big|_c.
     \label{rho_bc}
\end{multline}

The statistics of outcomes of joint measurements on the emitter and the fluorescence can be estimated from Eq.~\eqref{rho_bc}, and we proceed to do that shortly. However, we note that additional information is also contained in the state of fluorescence alone $\rho_c$, which can be accessed using complementary measurement strategies. For example, the probability distribution of fluorescent counts is given by,
\begin{multline}
    P_n = \bra{n}\rho_c\ket{n} = \left\{\frac{\sqrt{\gamma_0 \gamma_s}}{\gamma} \left[\text{cos}\left(\sqrt{\Delta t \gamma}\right)  - 1\right]\right\}^{2n} \\ \times\int d^2\alpha P(\alpha) \frac{|\alpha|^{2n}}{n!}e^{-\frac{\gamma_0 \gamma_s}{\gamma^2} \left[\text{cos}\left(\sqrt{\Delta t \gamma}\right)  - 1 \right]^2|\alpha  |^2},
     \label{P_nmain}
\end{multline}
and the resultant counting statistics of fluorescence is described in Appendix~\ref{appendix_Counting_stats}. Similarly, the statistics of measurements on the emitter alone would be equivalent to the results presented in Refs.~\cite{manikandan_testing_2025,manikandan2025complementary,manikandan2025probing,manikandan2026detector}.

\section{Quantum aspects of the drive via joint emitter-fluorescence measurements}
\label{section_probing}
We first consider joint quadrature measurements of the emitter and its fluorescent field. Our goal is to show that such correlation measurements reveal the quantum noise of the driving field relative to that of a coherent state of the driving field. 
The $P$-representation is quite helpful in this regard, as one can determine the quantum noise matrix (the covariance matrix) for the position and momentum quadratures of the driving radiation field as (also see Appendix~\ref{Appendix_Cov_mat_rad_filed}),
    \begin{equation}
    \begin{gathered}
       \begin{bmatrix}
\mathrm{Var}(\hat{x})_a & \mathrm{Cov}(\hat{x}, \hat{p})_a \\
\mathrm{Cov}(\hat{p}, \hat{x})_a & \mathrm{Var}(\hat{p})_a  \\
\end{bmatrix}= \\
       \begin{bmatrix}
        2\text{Var}(\Re(\alpha)) + \frac{1}{2} & 2\text{Cov}(\Re(\alpha), \Im(\alpha)) \\
       2\text{Cov}(\Re(\alpha), \Im(\alpha)) & 2\text{Var}(\Im(\alpha)) + \frac{1}{2} \\
       \end{bmatrix} ,
      \end{gathered}
      \label{Cov_mat_a_mode}
\end{equation}
where $\Re(\alpha)$, $\Im(\alpha)$ are the real and imaginary parts of $\alpha$ respectively, and the covariances in the right hand side are evaluated with respect to the Sudarshan-Glauber $P$-distribution function, $P(\alpha)$.

Performing simultaneous measurements of position and/or momentum quadratures on the emitter and its fluorescent field, denoted as $\{\hat{x}_b, \hat{p}_b, \hat{x}_c, \hat{p}_c \}$, gives the probability distributions of the outcomes (see Appendix~\ref{Appendix_Cov_mat_sim_meas}),
\begin{eqnarray}
    P_{\ell,j} &=& {}_{b}\langle \ell | \otimes {}_{c}\langle j| \rho_{bc} | \ell \rangle_b \otimes\ket{j}_c\nonumber\\& =& \frac{1}{\pi}\int d^2\alpha P(\alpha) e^{-s(\ell,\alpha, \Delta t, \gamma)} e^{-q(j,\alpha, \Delta t, \gamma)},
    \label{Plj}
\end{eqnarray}
where,
\begin{eqnarray}
   &&s(\ell,\alpha, \Delta t, \gamma) = \nonumber\\&&\begin{cases} 
\left(x_b-\Im(\alpha) \sqrt{\frac{2\gamma_0}{\gamma}} \text{sin}\left(\sqrt{\Delta t \gamma }\right)\right)^2, & \ell = x \\ 
\left(p_b + \Re(\alpha) \sqrt{\frac{2\gamma_0}{\gamma}} \text{sin}\left(\sqrt{\Delta t \gamma }\right)\right)^2, & \ell = p 
\end{cases}\\\nonumber
\end{eqnarray}
and
\begin{eqnarray}
  &&  q(j,\alpha, \Delta t, \gamma) =\nonumber\\
&&\begin{cases} 
\left(x_c-\Re(\alpha) \frac{\sqrt{2\gamma_0 \gamma_s}}{\gamma} \left[\text{cos}\left(\sqrt{\Delta t \gamma}\right)  - 1 \right]\right)^2, & j = x \\ 
\left( p_c - \Im(\alpha)\frac{\sqrt{2\gamma_0 \gamma_s}}{\gamma} \left[\text{cos}\left(\sqrt{\Delta t \gamma}\right)  - 1 \right] \right)^2, & j = p.
\end{cases}\\
\end{eqnarray}

These results allow to estimate the covariances of the position and momentum quadrature measurements on the emitter and its fluorescence as,
\begin{eqnarray}
    \braket{\hat{\ell}_b \hat{j_c}} - \braket{\hat{\ell}_b}\braket{\hat{j_c}} &=& 
    \int \ell_b j_c  P_{\ell,j} d\ell_c dj_b \nonumber\\&-& \int \ell_b P_{\ell,j} d\ell_b dj_c \int j_c P_{\ell,j}  d\ell_b dj_c,
\end{eqnarray}
for $\hat{\ell} = \{\hat{x},\hat{p}\}$ and $\hat{j} = \{\hat{x},\hat{p}\}$.
Performing the integration over all possible combinations of position and momentum quadratures (see Appendix~\ref{Appendix_Cov_mat_sim_meas}) the results can be summarized as
\begin{equation}
\begin{aligned}
 & {}\braket{\hat{\ell}_b \hat{j_c}} - \braket{\hat{\ell}_b}\braket{\hat{j_c}} = \\
&\quad F(\gamma_0, \gamma_s, \Delta t) \times
\begin{cases} 
\text{Var}\left(\Im(\alpha)\right), & \ell=x,\ j=p \\ 
-\text{Var}\left(\Re(\alpha)\right), & \ell=p,\ j=x \\ 
\text{Cov}\left(\Re(\alpha),\Im(\alpha)\right), & \ell=j=x \\
-\text{Cov}\left(\Re(\alpha),\Im(\alpha)\right), & \ell=j=p,
\end{cases}
\end{aligned}
\label{joint_meas_cov}
\end{equation}
where $F(\gamma_0, \gamma_s, \Delta t) \equiv \frac{2\gamma_0 \sqrt{\gamma_s}}{\sqrt{\gamma}\gamma} \text{sin}(\sqrt{\Delta t \gamma }) \left[\text{cos}(\sqrt{\Delta t \gamma})  - 1 \right] $.

Combining the results of Eq.~\eqref{joint_meas_cov} and Eq.~\eqref{Cov_mat_a_mode} allows to relate the outcomes of joint measurements of the emitter and its fluorescence to the covariance matrix of the joint incompatible measurements of position and momentum quadratures of the driving field as,
\begin{widetext}
    \begin{equation}
    \begin{gathered}
           \begin{bmatrix}
\braket{\hat{p}_b \hat{x}_c} - \braket{\hat{p}_b}\braket{\hat{x}_c} & \braket{\hat{p}_b \hat{p}_c} - \braket{\hat{p}_b}\braket{\hat{p}_c} \\
\braket{\hat{x}_b \hat{x}_c} - \braket{\hat{x}_b}\braket{\hat{x}_c} & \braket{\hat{x}_b \hat{p}_c} - \braket{\hat{x}_b}\braket{\hat{p}_c}  \\
\end{bmatrix} =
   \frac{\gamma_0 \sqrt{\gamma_s}}{\sqrt{\gamma}\gamma} \text{sin}(\sqrt{\Delta t \gamma }) \left[\text{cos}(\sqrt{\Delta t \gamma})  - 1 \right] \begin{bmatrix}
-\mathrm{Var}(\hat{x})_a + \frac{1}{2} & -\mathrm{Cov}(\hat{x}, \hat{p})_a \\
\mathrm{Cov}(\hat{p}, \hat{x})_a & \mathrm{Var}(\hat{p})_a - \frac{1}{2} \\
       \end{bmatrix}.
      \end{gathered}%
\label{joint_meas_bc_covar_a}
\end{equation}
\end{widetext}
Importantly, 
we see from Eq.~\eqref{joint_meas_bc_covar_a} that the correlations measure the quantum covariance matrix relative to the maximally classical coherent state of the driving field, in that the measured correlations vanish when the driving field is in a coherent state for which $\mathrm{Var}(\hat{x})_a=\mathrm{Var}(\hat{p})_a=1/2$ and $\mathrm{Cov}(\hat{p}, \hat{x})_a = 0$. 
Hence the correlations between the quantum emitter and its fluorescence probe the truly quantum mechanical features of the driving field in terms of the performed measurements.
Our result is valid at short times, in that we consider statistics derived from a single shot joint measurement of the quantum emitter and its fluorescence. 
At short times the prefactor becomes up to fifth order $F(\gamma_0, \gamma_s, \Delta t) \approx  -\frac{\gamma_0 \sqrt{\gamma_s}}{2} \left[(\sqrt{\Delta t })^3 - \frac{1}{4} (\sqrt{\Delta t })^5  \gamma \right]$.
Generically, our findings demonstrate that joint position-momentum quadrature measurements on the emitter and its fluorescence are able to probe the truly quantum features of the driving field.
Gaussian states are fully characterized by their quantum noise (covariance) matrix, which in its most general form---corresponding to squeezed, displaced, thermal states of the radiation field---has the following well-known expression
    \begin{equation}
    \begin{gathered}
    \mathcal{V}_\mathrm{G} = \frac{2 \bar{n}_{\mathrm{th}} + 1}{2} \times \\
           \begin{bmatrix}
\cosh(2r) - \cos(\phi) \sinh(2r)  & -\sin(\phi) \sinh(2r) \\
-\sin(\phi) \sinh(2r) & \cosh(2r) + \cos(\phi) \sinh(2r)  \\
\end{bmatrix},
      \end{gathered}%
\label{V_gauss}
\end{equation}
where $\bar{n}_{\mathrm{th}}$ denotes the average occupation number of the thermal state, so that the prefactor $(2\bar{n}_{\mathrm{th}}+1)/2$ represents uniformly distributed thermal noise, while the terms containing $r$ and $\phi$ describe the squeezing of the state.
Our approach therefore also enables a complete reconstruction of $\mathcal{V}_{G}$ while treating the coherent driving field as a reference, serving as a null test of classicality. 
An intrinsically quantum Gaussian state, such as squeezed state, is characterized by  $\mathrm{Var}(\hat{x})_a < 1/2$ or $\mathrm{Var}(\hat{p})_a < 1/2$ \cite{gerry2006introductory}, and would induce non-zero entries in the diagonal of the matrix in Eq.~\eqref{joint_meas_bc_covar_a}.

As for a coherent driving field each part of the system in Eq.~\eqref{exact_results} is in a product state of coherent fields at all times, another approach to observe the onset of non-classical features for a Gaussian driving field is through its purity, which can be calculated as \cite{adesso2004extremal} 
    $\mathcal{P}(\rho_a) = 1/\sqrt{\text{det}\left[2\mathcal{V}_{\mathrm{G}}\left(\rho_a\right)\right]}$.
The value $ \mathcal{P}(\rho)\leq 1$ indicates that the state is mixed, and consequently the drive is not in a coherent state.

A comparable null test for classicality of the driving field is the covariance of counts between the emitter and its fluorescence, as outlined in Appendix~\ref{appendix_Counting_stats}. 
From Eq.~\eqref{rho_bc} we obtain that,
\begin{eqnarray}
    &&\langle {\hat{n}}_b {\hat{n}}_c\rangle -    \langle {\hat{n}}_b\rangle\langle {\hat{n}}_c\rangle = \frac{\gamma_0^2\gamma_s}{\gamma^3}\sin^2\left(\sqrt{\Delta t\gamma}\right)\nonumber\\&\times&\left[\text{cos}\big(\sqrt{\Delta t \gamma}\big)  - 1 \right]^2\left(g^{(2)}_{\rho_a}(0)- 1 \right)\langle {\hat{n}}_a\rangle^2,
\end{eqnarray}
where $\langle{\hat{n}}_a\rangle=\braket{\hat{a}^\dagger \hat{a}}_{\rho_a}$ and  $g^{(2)}_{\rho_a}(0) \equiv  \frac{\braket{(\hat{a}^\dagger)^2 \hat{a}^2}_{\rho_a}}{\braket{\hat{a}^\dagger \hat{a}}_{\rho_a}^2}$ is the second order coherence function for the driving field.
Given that $g^{(2)}_{\rho_a}(0) = 1$ for a coherent state and $g^{(2)}_{\rho_a}(0) = 1+\frac{Q}{\langle {\hat{n}}_a\rangle}$ for generically different other states, where $Q$ is the Mandel's $Q$ parameter for the driving field, a non-zero value of the covariance of counts is another indication of the deviation from a coherent state of the driving field.

Measurements of fluorescence counts alone, described by Eq.~\eqref{P_nmain} also has interesting quantum sensing prospects. For instance, as outlined further in Appendix~\ref{appendix_Counting_stats}, the counting statistics of resonance fluorescence can in principle probe the second order coherence properties of the driving field in terms of the observed variance,
\begin{multline}
    (\Delta \hat{n}_c)^2  = \left[\frac{\braket{\hat{n}_c(\hat{n}_c-1)}}{\braket{\hat{n}_c}^2} - 1 \right] \braket{\hat{n}_c}^2 + \braket{\hat{n}_c} = \\
     G(\Delta t, \gamma)^2 \left(g^{(2)}_{\rho_a}(0)- 1 \right)  \braket{\hat{n}_a}^2 +  G(\Delta t, \gamma) \braket{\hat{n}_a}
\end{multline}
where $G(\Delta t, \gamma) \equiv \left\{\frac{\sqrt{\gamma_0 \gamma_s}}{\gamma} \left[\text{cos}(\sqrt{\Delta t \gamma})  - 1\right]\right\}^{2}$. Despite not having the null test property, such indirect measurements of the counting statistics of radiation fields can be of particular interest in the large $\langle \hat{n}_a\rangle\gg 1$ limit of the incoming radiation field, where a small leakage $\gamma_s\ll 1$ can attenuate the signal substantially to probe its quantum characteristics at the level of few fluorescent quanta.

\section{Physical Applications}
\label{section_discussion}

We have demonstrated that joint quantum measurements on a quantum harmonic emitter and its resonance fluorescence field can serve as a gateway to study quantum properties of the radiation field that is driving the emitter. Our analytically tractable model reveals that one can in fact reconstruct the full quantum noise matrix (the covariance matrix) of the driving field in generic quantum mechanical states from probing the statistical correlations between the outcomes of these measurements. Gaussian states---squeezed, displaced, thermal states of the radiation field---are fully characterized by their quantum covariance matrix suggesting that the prescribed quantum sensing strategy provides theoretically complete information for Gaussian states~\cite{simon1994quantum}. Various information theoretic measures such as the purity and entropy~\cite{adesso2007entanglement} of the radiation field can be extracted from the quantum covariance matrix in a straightforward manner. 
 Importantly, the joint-measurements approach outlined in this paper serves as a null test of classicality, as it probes the quantum covariance matrix of the driving field relative to the maximally classical (coherent) state, and the observed correlations vanish when the driving field is in a coherent state. 

Our results are particularly relevant for quantum metrology and sensing, where the driven emitter in our problem can be understood as the detector for the driving field one intends to probe, where the emission or leakage from the detector is guided and monitored simultaneously as resonance fluorescence. For oscillating charged dipoles, the fluorescent emission rate $\gamma_s$ in vacuum, predicted by the standard Wigner-Weisskopf theory typically scales (in three dimensions) as $\gamma_s\sim \frac{\omega^3 |d|^2}{\epsilon_0\hbar c^3},$
where $d$ is the transition dipole moment of the emitter~\cite{milonni2013quantum,Weisskopf:1930au}. In the case of phononic emitters, fluorescence emission can be through matter waves, and such scenarios are of foundational interest since they probe the extensibility of principles of quantum optics to quantum acoustics~\cite{krinner2018spontaneous,stewart2017analysis,stewart2020dynamics}. 

Although engineering the joint measurements can be challenging, simple extensions of the models described in Refs.~\cite{krinner2018spontaneous,stewart2017analysis} are possible, which makes the physical scenario of interest experimentally realizable in the quantum acoustic framework. 
These earlier works have considered the emission of matter waves from a site of single occupancy  via an optically mediated coupling, where the site corresponds to the ground state of a quantum harmonic trap. The scenario considered in Refs.~\cite{krinner2018spontaneous,stewart2017analysis}  can be extended to the coherently populated ground state of the harmonic trap (such as the case for a Bose-Einstein condensate in the trap), which will also spontaneously emit matter-wave quanta.  The key difference to optical emission is that here the continuum matter waves obey a quadratic dispersion relation. Simple emission scenarios consider a one dimensional ``wave-guide", and the emission can occur at the rate (within the Markov approximation)~\cite{krinner2018spontaneous,stewart2017analysis},
\begin{eqnarray}
    \gamma_s = \sqrt{\frac{\pi\Omega^4}{2\omega_0\Delta}}e^{-2\Delta/\omega_0}.
\end{eqnarray}
Above, $\Omega$ is the effective coupling rate to the optical field that mediates the population transfer and $\Delta$ is the energy difference between this optical field and the difference of bare energies of the trap and the one dimensional waveguide.
In the resonant approximation the emitted matter wave will have the wave-number and energy determined by the relation $\hbar^2k^2/(2m)=\hbar\Delta$. When the population in the harmonic trap is stimulated or driven externally, the emission can be understood as resonance fluorescence.

Such acoustic sources and detectors are also of interest in the quantum gravitational context where they have been proposed as capable detectors for recording single graviton exchange events~\cite{tobar2024detecting,shenderov2026stimulated} and to probe their statistics~\cite{manikandan_testing_2025,manikandan2025complementary,manikandan2026detector,manikandan2025probing,toccacelo2026quantum}. Here, statistical tests based on measurements on one or more detectors have been shown to reveal the quantum properties of the radiation field as tests of the coherent state hypothesis~\cite{manikandan_testing_2025,manikandan2025complementary,manikandan2026detector,manikandan2025probing}. 
Unlike an oscillating charged dipole relevant in the electromagnetic context, the fundamental oscillator relevant in the gravitational context is a mass quadrupole~\cite{maggiore2008gravitational}, with the rate $\gamma_0=8GML^2\omega^4/(\pi^4c^5)$~\cite{tobar2024detecting,manikandan_testing_2025}, where $G$ is the Newton's constant, $M$ is the acoustic resonator's mass, and $L$ its length. For reasonable masses and for $\omega\sim$ kHz frequencies, one obtains $\gamma_0 \sim 10^{-33}$ Hz, which is tiny; however the detection of gravitons, and probing their quantum statistics are still expected to be possible given that the incoming gravitational radiation field that LIGO is sensitive to, contains as many as $10^{36}$ gravitons, which compensates for the small $\gamma_0$~\cite{manikandan_testing_2025,manikandan2025complementary,manikandan2026detector,manikandan2025probing}. 
In this case as well, our findings have an interesting implication that joint measurements on quadrupole oscillator and the emitted fluorescent radiation could in principle provide additional tests for probing the non-classicality of the gravitational field. Such tests can be complementary to the statistical null tests using multiple quadrupole detectors proposed in Ref.~\cite{manikandan2026detector}.

In broader terms, our findings suggest that leaky resonators can also be informative, provided the leakage is guided and monitored, as in resonance fluorescence~\cite{lewalle2020measuring,jordan2016anatomy,ficheux2018dynamics}. Higher order correlations of the driving field may in principle also be extracted systematically  using the methodologies we presented, and they could provide valuable insights on characteristically quantum features of the driving field, beyond the second moment. Similarly, the complete covariance matrix for measurements in the emitter-fluorescence sector can probe the quantum entanglement induced by the driving field between the emitter and its fluorescence as well~\cite{simon1994quantum}. We defer a systematic characterization of these to future work. 
\textit{Author contributions}---The work was conceptualized by SKM. Both YB and SKM contributed equally to the calculations, analyzing the results, and to writing the manuscript.

\textit{Acknowledgments}---SKM acknowledges the support from the Department of Atomic Energy, Government of India, under Project Identification No. RTI4007.  The early discussions that led to this collaboration were initiated at the Nordic Institute for Theoretical Physics (Nordita), Stockholm, Sweden. SKM acknowledges helpful discussions on related topics with Frank Wilczek. YB acknowledges Jens Bardarson for his advice. YB received funding from the European Research Council (ERC) under
the European Union's Horizon 2020 research and innovation program (Grant Agreement No. 101001902).

\appendix

\section{Solution via the normal modes}
\label{appendix_normal_modes_solution}
We start by rewriting Eq.~\eqref{Time_evo_U_I} as 
 \begin{multline}
     \hat{U}_{I}\ket{\alpha}_a\ket{0}_b \ket{0}_c = \\
     = e^{-i \sqrt{\Delta t}\left[(\sqrt{\gamma_0}\hat{a}^\dagger  +  \sqrt{\gamma_s} \hat{c}^\dagger)\hat{b} + (\sqrt{\gamma_0}\hat{a}  + \sqrt{\gamma_s} \hat{c} )\hat{b}^\dagger\right]}\ket{\alpha}_a\ket{0}_b \ket{0}_c.
 \end{multline}
We can define the new creation and annihilation operators as
\begin{equation}
\hat{f}^{\dagger}_+ \equiv \frac{1}{\sqrt{\gamma}}(\sqrt{\gamma_0}\hat{a}^\dagger  +  \sqrt{\gamma_s} \hat{c}^\dagger),
\end{equation}
\begin{equation}
\hat{f}_+ \equiv \frac{1}{\sqrt{\gamma}}(\sqrt{\gamma_0}\hat{a}  +  \sqrt{\gamma_s} \hat{c}),
\end{equation}
which is the first normal mode of the system. Above, $\gamma \equiv \gamma_0+\gamma_s.$
The second normal mode of the system is defined as
\begin{equation}
\hat{f}^{\dagger}_- \equiv \frac{1}{\sqrt{\gamma}}(\sqrt{\gamma_s}\hat{a}^\dagger  - \sqrt{\gamma_0}\hat{c}^\dagger),
\end{equation}
\begin{equation}
\hat{f}_- \equiv  \frac{1}{\sqrt{\gamma}}(\sqrt{\gamma_s}\hat{a} - \sqrt{\gamma_0}\hat{c}).
\end{equation}
The modes are orthonormal and fulfill bosonic commutation relations $ [\hat{f}_+, \hat{f}^{\dagger}_+] =  [\hat{f}_-, \hat{f}^{\dagger}_-] = 1$ and $[\hat{f}_+, \hat{f}_+] = [\hat{f}^{\dagger}_+, \hat{f}^{\dagger}_+] = [\hat{f}^{\dagger}_-, \hat{f}^{\dagger}_-] = [\hat{f}_-, \hat{f}_-] = 0$, while they also commute with each other $[\hat{f}_-, \hat{f}^{\dagger}_+] = [\hat{f}^{\dagger}_-, \hat{f}_+] = 0$.

The unitary time evolution in terms of the two normal modes becomes
 \begin{equation}
     \hat{U}_{I}\ket{\alpha}_a\ket{0}_b \ket{0}_c = e^{-i \sqrt{\Delta t\gamma}(\hat{f}^{\dagger}_+ \hat{b} + \hat{f}_+ \hat{b}^\dagger)} \ket{+} \ket{0}_b \ket{-},
     \label{normal_modes_evo}
 \end{equation}
where the $\ket{+}$ and $\ket{-}$ are the two normal modes in coherent states defined as
 \begin{equation}
     \ket{+} \equiv \Big|\alpha\sqrt{\frac{\gamma_0}{\gamma}}\Big\rangle_+
 \end{equation}
 and
  \begin{equation}
     \ket{-} \equiv \Big|\alpha\sqrt{\frac{\gamma_s}{\gamma}}\Big\rangle_-,
 \end{equation}
 such that $\hat{f}_+\ket{+}=\alpha\sqrt{\frac{\gamma_0}{\gamma}}\ket{+} =  \alpha\sqrt{\frac{\gamma_0}{\gamma}} \ket{\alpha}_a \ket{0}_c$ and
$\hat{f}_-\ket{-}=\alpha\sqrt{\frac{\gamma_s}{\gamma}}\ket{-} =  \alpha\sqrt{\frac{\gamma_s}{\gamma}} \ket{\alpha}_a \ket{0}_c$.

 We now note that (see for example, ref.~\cite{manikandan_testing_2025}),
 \begin{eqnarray}
     &&e^{-i \sqrt{\Delta t\gamma}(\hat{f}^{\dagger}_+ \hat{b} + \hat{f}_+ \hat{b}^\dagger)} \ket{+} \ket{0}_b \nonumber\\&=& \Big|\alpha\sqrt{\frac{\gamma_0}{\gamma}}\cos(\sqrt{\gamma\Delta t})\Big\rangle_+ \otimes \Big|-i\alpha\sqrt{\frac{\gamma_0}{\gamma}}\sin(\sqrt{\gamma\Delta t})\Big\rangle_b.\nonumber\\
 \end{eqnarray}
 Rewriting the normal modes in coherent states as displaced vacuum states, we can write Eq.~\eqref{normal_modes_evo} as
\begin{multline}
     \hat{U}_{I}\ket{\alpha}_a\ket{0}_c \ket{0}_b = \\
    e^{-|\alpha|^2/2} 
    e^{\alpha\sqrt\frac{\gamma_0}{\gamma}\text{cos}(\sqrt{\Delta t \gamma}) \hat{f}^{\dagger}_+} 
    e^{- i \alpha\sqrt{\frac{\gamma_0}{\gamma}}\text{sin}(\sqrt{\Delta t \gamma}) \hat{b}^{\dagger}} \\
    e^{\alpha \sqrt\frac{\gamma_s}{\gamma} \hat{f}^{\dagger}_-}\ket{0}_+ \ket{0}_b \ket{0}_- =  \\
    = e^{-|\alpha|^2/2}  e^{\alpha (\text{cos}(\sqrt{\Delta t \gamma})\frac{\gamma_0}{\gamma} + \frac{\gamma_s}{\gamma})\hat{a}^\dagger} e^{- i \alpha \sqrt{\frac{\gamma_0}{\gamma}}\text{sin}(\sqrt{\Delta t \gamma}) \hat{b}^{\dagger}} \\
     e^{ \alpha (\text{cos}(\sqrt{\Delta t \gamma})\frac{\sqrt{\gamma_0 \gamma_s}}{\gamma}  - \frac{\sqrt{\gamma_0 \gamma_s}}{\gamma})\hat{c}^\dagger}\ket{0}_a \ket{0}_b \ket{0}_c = \\
    =\left|\alpha \frac{1}{\gamma}(\gamma_0 \text{cos}(\sqrt{\Delta t \gamma}) + \gamma_s )\right\rangle_a \otimes 
     \left|-\alpha  i \sqrt{\frac{\gamma_0}{\gamma}} \text{sin}(\sqrt{\Delta t \gamma })\right\rangle_b \otimes \\
     \left|\alpha  \frac{\sqrt{\gamma_0 \gamma_s}}{\gamma} \left(\text{cos}(\sqrt{\Delta t \gamma})  - 1 \right)\right\rangle_c,
\end{multline}
which is the same result as obtained in the main text via the Hadamard's lemma.
\section{Counting statistics}
\label{appendix_Counting_stats}
The covariance of counts between the emitter and its fluorescence can be calculated as 
\begin{multline}
\langle {\hat{n}}_b {\hat{n}}_c\rangle -    \langle {\hat{n}}_b\rangle\langle {\hat{n}}_c\rangle = \\
\sum_{n_b,n_c} n_b n_c P_{n_b,n_c} - \sum_{n_b,n_c} n_b P_{n_b,n_c} \sum_{n_b,n_c} n_c P_{n_b,n_c},
\end{multline}
where the probability of observing $n_b$ counts on the emitter and $n_c$ counts on the fluorescence using Eq.~\ref{rho_bc} is
\begin{multline}
P_{n_b,n_c} = \bra{n_b}\bra{n_c}\rho_{bc}\ket{n_b}\ket{n_c} = 
  \int d^2\alpha P(\alpha) \times \\ e^{-|-\alpha  i \sqrt{\frac{\gamma_0}{\gamma}}  \text{sin}(\sqrt{\Delta t \gamma })|^2} \frac{|- i \alpha \sqrt{\frac{\gamma_0}{\gamma}} \text{sin}(\sqrt{\Delta t \gamma })|^{2n_b} }{n_b!} \times \\ e^{-|\alpha  \frac{\sqrt{\gamma_0 \gamma_s}}{\gamma} (\text{cos}(\sqrt{\Delta t \gamma})  - 1 )|^2}  \frac{| \alpha \frac{\sqrt{\gamma_0 \gamma_s}}{\gamma} (\text{cos}(\sqrt{\Delta t \gamma})  - 1 )|^{2n_c} }{n_c!}
\end{multline}
Using the fact that $\int d^2\alpha P(\alpha) |\alpha|^2 = \int d^2\alpha P(\alpha) \bra{\alpha}\hat{a}^\dagger \hat{a} \ket{\alpha} = \braket{\hat{a}^\dagger \hat{a}}_{\rho_a}$,  $\int d^2\alpha P(\alpha) |\alpha|^4 = \int d^2\alpha P(\alpha) \bra{\alpha}(\hat{a}^\dagger)^2 \hat{a}^2 \ket{\alpha} = \braket{(\hat{a}^\dagger)^2 \hat{a}^2}_{\rho_a}$ and rewriting the sum as exponential the resulting expression is
\begin{eqnarray}
    &&\langle {\hat{n}}_b {\hat{n}}_c\rangle -    \langle {\hat{n}}_b\rangle\langle {\hat{n}}_c\rangle = \frac{\gamma_0^2\gamma_s}{\gamma^3}\sin^2\left(\sqrt{\Delta t\gamma}\right)\nonumber\\&\times&\left[\text{cos}\big(\sqrt{\Delta t \gamma}\big)  - 1 \right]^2\left(g^{(2)}_{\rho_a}(0)- 1 \right)\langle {\hat{n}}_a\rangle^2.
\end{eqnarray}

For the counting statistics of the resonance fluorescence the reduced density matrix of the fluorescence field is obtained by tracing out the $\hat{a}$- and $\hat{b}$- modes in Eq.~\eqref{exact_results_P_rep} 
\begin{multline}
    \rho_{c} =  \int d^2\alpha P(\alpha) \\
     |\alpha  \frac{\sqrt{\gamma_0 \gamma_s}}{\gamma} (\text{cos}(\sqrt{\Delta t \gamma})  - 1 )\rangle \langle\alpha  \frac{\sqrt{\gamma_0 \gamma_s}}{\gamma} (\text{cos}(\sqrt{\Delta t \gamma})  - 1 )|_c.
     \label{rho_c}
\end{multline}
Probability of the resonant fluorescence to be in state $\ket{n_c}$ is given by 
\begin{multline}
    P_{n_c} = \bra{n_c}\rho_c\ket{n_c} = \left(\frac{\sqrt{\gamma_0 \gamma_s}}{\gamma} (\text{cos}(\sqrt{\Delta t \gamma})  - 1)\right)^{2n_c} \\ \int d^2\alpha P(\alpha) \frac{|\alpha|^{2n_c}}{n_c!}e^{-\frac{\gamma_0 \gamma_s}{\gamma^2} (\text{cos}(\sqrt{\Delta t \gamma})  - 1 )^2|\alpha  |^2}.
     \label{P_n}
\end{multline}
From this, the mean number of fluorescent quanta is calculated as
\begin{multline}
    \braket{\hat{n}_c} = \sum_{n_c}^\infty n_c P_{n_c} = \\
       \left\{\frac{\sqrt{\gamma_0 \gamma_s}}{\gamma} \left[\text{cos}(\sqrt{\Delta t \gamma})  - 1\right]\right\}^{2} \braket{\hat{a}^\dagger \hat{a}}_{\rho_a}.
\end{multline}
The second moment can be calculated as
\begin{equation}
    \braket{\hat{n}_c^2} =  \braket{\hat{n}_c(\hat{n}_c-1) + \hat{n}_c} =  \braket{\hat{n}_c(\hat{n}_c-1)} + \braket{\hat{n}_c}.
\end{equation}
The first term gives
\begin{multline}
     \braket{\hat{n}_c(\hat{n}_c-1)} = \sum_{n_c}^\infty n_c (n_c-1) P_{n_c} = \\
      \left[\frac{\sqrt{\gamma_0 \gamma_s}}{\gamma} (\text{cos}(\sqrt{\Delta t \gamma})  - 1)\right]^{4} \braket{(\hat{a}^\dagger)^2 \hat{a}^2}_{\rho_a}.
\end{multline}
Combining these results we can obtain the variance of the counting statistics of the fluorescence
\begin{multline}
    (\Delta \hat{n}_c)^2  = \left[\frac{\braket{\hat{n}_c(\hat{n}_c-1)}}{\braket{\hat{n}_c}^2} - 1 \right] \braket{\hat{n}_c}^2 + \braket{\hat{n}_c} = \\
     G(\Delta t, \gamma) \left[\left(g^{(2)}_{\rho}(0)- 1 \right) G(\Delta t, \gamma)  \braket{\hat{a}^\dagger \hat{a}}_{\rho_a}^2 +  \braket{\hat{a}^\dagger \hat{a}}_{\rho_a} \right]
\end{multline}
where $G(\Delta t, \gamma) \equiv \left\{\frac{\sqrt{\gamma_0 \gamma_s}}{\gamma} \left[\text{cos}(\sqrt{\Delta t \gamma})  - 1\right]\right\}^{2}$ and we have observed that  
\begin{equation}
    \frac{\braket{(\hat{a}^\dagger)^2 \hat{a}^2}_{\rho_a}}{\braket{\hat{a}^\dagger \hat{a}}_{\rho_a}^2} = g^{(2)}_{\rho}(0),
\end{equation}
is the second order coherence function for the driving field in quantum state $\rho_a$.

\section{Covariance matrix of the radiation field}
\label{Appendix_Cov_mat_rad_filed}
The covariance matrix of the radiation field is defined as

\begin{equation}
\begin{gathered}
\begin{bmatrix}
\mathrm{Var}(\hat{x})_a & \mathrm{Cov}(\hat{x}, \hat{p})_a \\
\mathrm{Cov}(\hat{p}, \hat{x})_a & \mathrm{Var}(\hat{p})_a  \\
\end{bmatrix} \\
=\begin{bmatrix}
\braket{\hat{x}_a^2} - \braket{\hat{x}_a}^2 & \frac{\braket{\hat{x}_a \hat{p}_a+\hat{p}_a \hat{x}_a}}{2} - \braket{\hat{x}_a}\braket{\hat{p}_a} \\
\frac{\braket{\hat{x}_a \hat{p}_a+\hat{p}_a \hat{x}_a}}{2} - \braket{\hat{x}_a}\braket{\hat{p}_a} & \braket{\hat{p}_a^2} - \braket{\hat{p}_a}^2  \\
\end{bmatrix}
\label{Cov_matrix_drive}
\end{gathered}
\end{equation}
The expectation values for the position quadrature, squared-position quadrature, momentum quadrature, and squared-momentum quadrature are 
\begin{equation}
    \braket{ \hat{x} } =  1/\sqrt{2} (\braket{\alpha | a | \alpha} + \braket{\alpha | a^\dagger | \alpha}) =  \sqrt{2} \Re(\alpha),
    \label{exp_x_a}
\end{equation}
\begin{multline}
    \braket{ \hat{x}^2 } =    
    1/2 (\alpha^2 + 2|\alpha|^2 + (\alpha^*)^2 +1) = 2 \Re^2(\alpha) + 1/2,
    \label{exp_x_a2}
\end{multline}
\begin{equation}
    \braket{\hat{p}} =  \frac{1}{i\sqrt{2}} (\alpha - \alpha^*) = \sqrt{2} \Im(\alpha),
    \label{exp_p_a}
\end{equation}
and
\begin{multline}
    \braket{\hat{p}^2} =  
    -1/2 (\alpha^2 - 2|\alpha|^2 + (\alpha^*)^2 -1) =  2 \Im^2(\alpha) + 1/2
    \label{exp_p_a2}
\end{multline}
respectively, where we have used the definitions of position and momentum quadratures of the harmonic oscillator $\hat{x} = 1/\sqrt{2} (a + a^\dagger)$,
$\hat{p} = 1/i\sqrt{2} (a - a^\dagger)$
and properties of complex numbers
$\alpha + \alpha^* = 2\Re(\alpha)$,
$\alpha - \alpha^* = 2i\Im(\alpha)$.
For the off-diagonal covariance terms we also calculate the following term
\begin{multline}
    \braket{\hat{x} \hat{p} + \hat{p} \hat{x} }  =\frac{1}{i}(\alpha^2 - (\alpha^*)^2) = 4\Re(\alpha)\Im(\alpha).
    \label{exp_px_a_xp_a}
\end{multline}
Using the results of Eqs.~\eqref{exp_x_a}-\eqref{exp_px_a_xp_a} in Eq.~\eqref{Cov_matrix_drive} the final expression becomes

    \begin{equation}
    \begin{gathered}
       \begin{bmatrix}
\mathrm{Var}(\hat{x})_a & \mathrm{Cov}(\hat{x}, \hat{p})_a \\
\mathrm{Cov}(\hat{p}, \hat{x})_a & \mathrm{Var}(\hat{p})_a  \\
\end{bmatrix}=\\
       \begin{bmatrix}
        2\text{Var}(\Re(\alpha)) + \frac{1}{2} & 2\text{Cov}(\Re(\alpha), \Im(\alpha)) \\
       2\text{Cov}(\Re(\alpha), \Im(\alpha)) & 2\text{Var}(\Im(\alpha)) + \frac{1}{2} \\
       \end{bmatrix}.
      \end{gathered}%
      \label{Cov_mat_a_mode_app}
\end{equation}

\section{Joint measurements of the position-momentum quadratures}
\label{Appendix_Cov_mat_sim_meas}
Using the overlap of the position and momentum states with the coherent state $\braket{x|\alpha} = \pi^{-1/4}  e^{-1/2(x-\sqrt{2}\Re(\alpha))^2 + ix \sqrt{2}\Im(\alpha) -i\Re(\alpha)\Im(\alpha)}$ and $\braket{p|\alpha} = \pi^{-1/4} e^{- 1/2( p - \sqrt{2}\Im(\alpha))^2 - i  p  \sqrt{2}\Re(\alpha)  + i\Re(\alpha)\Im(\alpha)}$ on the reduced density matrix of the $\hat{b}$- and $\hat{c}$-modes in Eq.~\eqref{rho_bc} we get the following probability distributions of simultaneous measurements

\begin{multline}
    P_{x,p} = {}_{b}\langle x | \otimes {}_{c}\langle p| \rho_{bc} | x \rangle_b \otimes\ket{p}_c = \frac{1}{\pi}\int d^2\alpha P(\alpha) \\
    e^{-(x_b-\Im(\alpha) \sqrt{\frac{2\gamma_0}{\gamma}} \text{sin}(\sqrt{\Delta t \gamma }))^2}  e^{- ( p_c - \Im(\alpha)\frac{\sqrt{2\gamma_0 \gamma_s}}{\gamma} (\text{cos}(\sqrt{\Delta t \gamma})  - 1 ) )^2 },
    \label{Pxp}
\end{multline}

\begin{multline}
    P_{p,x} = {}_{b}\langle p | \otimes {}_{c}\langle x| \rho_{bc} | p \rangle_b \otimes\ket{x}_c =\frac{1}{\pi}\int d^2\alpha P(\alpha) \\ 
    e^{- ( p_b + \Re(\alpha) \sqrt{\frac{2\gamma_0}{\gamma}} \text{sin}(\sqrt{\Delta t \gamma }))^2 }  e^{-(x_c-\Re(\alpha) \frac{\sqrt{2\gamma_0 \gamma_s}}{\gamma} (\text{cos}(\sqrt{\Delta t \gamma})  - 1 ))^2},
    \label{Ppx}
\end{multline}

\begin{multline}
    P_{x,x} = {}_{b}\langle x | \otimes {}_{c}\langle x| \rho_{bc} | x \rangle_b \otimes\ket{x}_c = \frac{1}{\pi}\int d^2\alpha P(\alpha) \\
    e^{- ( x_b - \Im(\alpha)  \sqrt{\frac{2\gamma_0}{\gamma}} \text{sin}(\sqrt{\Delta t \gamma }))^2 }  e^{-(x_c- \Re(\alpha) \frac{\sqrt{2\gamma_0 \gamma_s}}{\gamma}  (\text{cos}(\sqrt{\Delta t \gamma})  - 1 ))^2},
    \label{Pxx}
\end{multline}    

\begin{multline}
    P_{p,p} = {}_{b}\langle p | \otimes {}_{c}\langle p| \rho_{bc} | p \rangle_b \otimes\ket{p}_c = \frac{1}{\pi}\int d^2\alpha P(\alpha) \\ 
    e^{- ( p_b + \Re(\alpha)  \sqrt{\frac{2\gamma_0}{\gamma}} \text{sin}(\sqrt{\Delta t \gamma }))^2 }  e^{-(p_c - \Im(\alpha)\frac{\sqrt{2\gamma_0 \gamma_s}}{\gamma} (\text{cos}(\sqrt{\Delta t \gamma})  - 1 ))^2}.
    \label{Ppp}
\end{multline}    
We can summarize the probabilities from Eqs.~\eqref{Pxp}-\eqref{Ppp} in a general expression
\begin{multline}
    P_{\ell,j} = {}_{b}\langle \ell | \otimes {}_{c}\langle j| \rho_{bc} | \ell \rangle_b \otimes\ket{j}_c = \\ \frac{1}{\pi}\int d^2\alpha P(\alpha) e^{-s(\ell,\alpha, \Delta t, \gamma)} e^{-q(j,\alpha, \Delta t, \gamma)},
    \label{append_Plj}
\end{multline} 
where
\[
s(\ell,\alpha, \Delta t, \gamma) =
\begin{cases} 
(x_b-\Im(\alpha) \sqrt{\frac{2\gamma_0}{\gamma}} \text{sin}(\sqrt{\Delta t \gamma }))^2, & \ell = x \\ 
(p_b + \Re(\alpha) \sqrt{\frac{2\gamma_0}{\gamma}} \text{sin}(\sqrt{\Delta t \gamma }))^2, & \ell = p 
\end{cases}
\]
and
\begin{eqnarray*}
    &&q(j,\alpha, \Delta t, \gamma) =\nonumber\\
&&\begin{cases} 
(x_c-\Re(\alpha) \frac{\sqrt{2\gamma_0 \gamma_s}}{\gamma} (\text{cos}(\sqrt{\Delta t \gamma})  - 1 ))^2, & j = x \\ 
( p_c - \Im(\alpha)\frac{\sqrt{2\gamma_0 \gamma_s}}{\gamma} (\text{cos}(\sqrt{\Delta t \gamma})  - 1 ) )^2, & j = p 
\end{cases}
\end{eqnarray*}
From Eqs.\eqref{Pxp}---\eqref{append_Plj}, we notice that when the driving radiation field is in a coherent state $|\alpha\rangle$, the probability distribution of the outcomes of position quadrature measurements on the emitter, $x_b$, and outcomes of momentum quadrature measurements on the fluorescence, $p_c$, only depend on the imaginary part of $\alpha$, which indicates their relation to the variance of the momentum quadrature of the driving field as can be seen in Eq.~\eqref{Cov_mat_a_mode} of the maintext.
Similarly the probability distribution of the outcomes of the measurements $p_b$ and $x_c$ depends only on the real part of $\alpha$, relating these measurements to the variance of the position quadrature of the driving field. The equivalent results for a generic quantum state of the driving field is captured by the corresponding $P$-distribution function averages as described by Eq.~\eqref{Cov_mat_a_mode} of the maintext. We obtain,
\begin{multline}
    \braket{p_b x_c} - \braket{p_b}\braket{x_c} = \\
    \int p_b x_c P_{p,x} dx_c dp_b - \int p_b P_{p,x} dx_c dp_b \int x_c P_{p,x} dx_c dp_b\\
    = - \frac{2\gamma_0 \sqrt{\gamma_s}}{\sqrt{\gamma}\gamma} \text{sin}(\sqrt{\Delta t \gamma }) (\text{cos}(\sqrt{\Delta t \gamma})  - 1 ) 
    \text{Var}(\Re(\alpha)).
\end{multline}
Similarly, position quadrature on the detector and momentum quadrature on the fluorescence yields the following correlation function,
\begin{multline}
    \braket{\hat{x}_b \hat{p}_c} - \braket{\hat{x}_b}\braket{\hat{p}_c} =  \\
    \int x_b p_c P_{x,p} dx_b dp_c - \int x_b P_{x,p} dx_b dp_c \int p_c P_{x,p} dx_b dp_c= \\
    \frac{2\gamma_0 \sqrt{\gamma_s}}{\sqrt{\gamma}\gamma} \text{sin}(\sqrt{\Delta t \gamma }) (\text{cos}(\sqrt{\Delta t \gamma})  - 1 ) \text{Var}(\Im(\alpha)).
\end{multline}
Position quadrature on the detector and position quadrature on the fluorescence yields,
\begin{multline}
    \braket{\hat{x}_b \hat{x}_c} - \braket{\hat{x}_b}\braket{\hat{x}_c} = \\
    \int x_b x_c P_{x,x} dx_b dx_c - \int x_b P_{x,x} dx_b dx_c \int x_c P_{x,x} dx_c dx_b = \\
    \frac{2\gamma_0 \sqrt{\gamma_s}}{\sqrt{\gamma}\gamma} \text{sin}(\sqrt{\Delta t \gamma }) (\text{cos}(\sqrt{\Delta t \gamma})  - 1 ) \text{Cov}(\Re(\alpha), \Im(\alpha)).
\end{multline}
And finally, momentum quadrature on the detector and momentum quadrature on the fluorescence yields the correlation function,
\begin{multline}
    \braket{\hat{p}_b \hat{p}_c} - \braket{\hat{p}_b}\braket{\hat{p}_c} = \\
     \int p_b p_c P_{p,p} dp_b dp_c  - \int p_b P_{p,p} dp_c dp_b \int p_c P_{p,p} dp_b dp_c = \\ 
     -\frac{2\gamma_0 \sqrt{\gamma_s}}{\sqrt{\gamma}\gamma} \text{sin}(\sqrt{\Delta t \gamma }) (\text{cos}(\sqrt{\Delta t \gamma})  - 1 ) \text{Cov}(\Im(\alpha), \Re(\alpha)).
\end{multline}

In the above equations, the variance and covariance of the real and imaginary parts of alpha are defined as,
$\text{Var}(\Re(\alpha)) = \int d^2\alpha \Re^2(\alpha) P(\alpha) - \Big(\int d^2\alpha \Re(\alpha) P(\alpha) \Big)^2$, 
$\text{Var}(\Im(\alpha)) = \int d^2\alpha \Im^2(\alpha) P(\alpha) - \Big(\int d^2\alpha \Im(\alpha) P(\alpha) \Big)^2$ and 
$\text{Cov}(\Re(\alpha), \Im(\alpha)) = \int d^2\alpha \Re(\alpha) \Im(\alpha) P(\alpha) - \int d^2\alpha \Re(\alpha) P(\alpha) \int d^2\beta \Im(\beta) P(\beta)$.

The outcome of correlation measurements in a matrix form is given by,
\begin{widetext}
    \begin{equation}
    \begin{gathered}
       \begin{bmatrix}
\braket{\hat{p}_b \hat{x}_c} - \braket{\hat{p}_b}\braket{\hat{x}_c} & \braket{\hat{p}_b \hat{p}_c} - \braket{\hat{p}_b}\braket{\hat{p}_c} \\
\braket{\hat{x}_b \hat{x}_c} - \braket{\hat{x}_b}\braket{\hat{x}_c} & \braket{\hat{x}_b \hat{p}_c} - \braket{\hat{x}_b}\braket{\hat{p}_c}  \\
\end{bmatrix}
= 
\frac{2\gamma_0 \sqrt{\gamma_s}}{\sqrt{\gamma}\gamma} \text{sin}(\sqrt{\Delta t \gamma }) (\text{cos}(\sqrt{\Delta t \gamma})  - 1 )
       \begin{bmatrix}
       -  \text{Var}(\Re(\alpha)) & -\text{Cov}(\Re(\alpha), \Im(\alpha)) \\
       \text{Cov}(\Re(\alpha), \Im(\alpha)) & \text{Var}(\Im(\alpha))  \\
       \end{bmatrix} .
       \\
       \label{Sim_meas_matrix}
      \end{gathered}%
\end{equation}
\end{widetext}

Combining the results of Eqs.~\eqref{Cov_mat_a_mode_app} and \eqref{Sim_meas_matrix} gives the covariance of simultaneous measurements on the detector and the fluorescence in terms of the variance and covariance of the radiation field,
\begin{widetext}
    \begin{equation}
    \begin{gathered}
           \begin{bmatrix}
\braket{\hat{p}_b \hat{x}_c} - \braket{\hat{p}_b}\braket{\hat{x}_c} & \braket{\hat{p}_b \hat{p}_c} - \braket{\hat{p}_b}\braket{\hat{p}_c} \\
\braket{\hat{x}_b \hat{x}_c} - \braket{\hat{x}_b}\braket{\hat{x}_c} & \braket{\hat{x}_b \hat{p}_c} - \braket{\hat{x}_b}\braket{\hat{p}_c}  \\
\end{bmatrix} = 
    \frac{\gamma_0 \sqrt{\gamma_s}}{\sqrt{\gamma}\gamma} \text{sin}(\sqrt{\Delta t \gamma }) (\text{cos}(\sqrt{\Delta t \gamma})  - 1 )
    \begin{bmatrix}
-\mathrm{Var}(\hat{x})_a + \frac{1}{2} & -\mathrm{Cov}(\hat{x}, \hat{p})_a \\
\mathrm{Cov}(\hat{p}, \hat{x})_a & \mathrm{Var}(\hat{p})_a - \frac{1}{2} \\
       \end{bmatrix}.
      \end{gathered}%
\label{Sim_meas_final_results}
\end{equation}
\end{widetext}
As highlighted in the main text, note here that the correlations probe the quantum noise (covariance) matrix of the driving field relative to a coherent state.
\bibliography{ref}
    \end{document}